# Shared urbanism: Big data on accommodation sharing in urban Australia

Somwrita Sarkar and Nicole Gurran

## Abstract

As affordability pressures and tight rental markets in global cities mount, online shared accommodation sites proliferate. Home sharing arrangements present dilemmas for planning that aims to improve health and safety standards, while supporting positives such as the usage of dormant stock and the relieving of rental pressures on middle/lower income earners. Currently, no formal data exists on this internationally growing trend. Here, we present a first quantitative glance on shared accommodation practices across all major urban centers of Australia enabled via collection and analysis of thousands of online listings. We examine, countrywide, the spatial and short time scale temporal characteristics of this market, along with preliminary analysis on rents, dwelling types and other characteristics. Findings have implications for housing policy makers and planning practitioners seeking to monitor and respond to housing policy and affordability pressures in formal and informal housing markets.

---

S. Sarkar (Corresponding author)
Architecture, Design and Planning, The University of Sydney, NSW 2006 Australia.
Email: somwrita.sarkar@sydney.edu.au

N. Gurran
Architecture, Design and Planning, The University of Sydney, NSW 2006 Australia.
Email: nicole.gurran@sydney.edu.au

## 1. Introduction

A city is a sum total of its physical-spatial structure and the social, technological, and economic processes that occur within this structure (Batty, 2013). While some aspects of these processes are planned, or regulated, most processes are self-organised, shaped by local peer-to-peer interactions. A particular example of self-organised processes can be observed in the rising phenomenon of the *shared economy* or what we term as *shared urbanism*: peer-to-peer sharing of resources enabling functions such as mobility (car sharing or pooling practices) and accommodation (flatsharing or housesharing). Such phenomenon present profound but largely unknown and unexplored implications for both mobility as well as housing policy and planning and urban regulation. They represent what may be termed as an "informal sector". The self-organised behavior of shared urbanism promises more adaptive, more responsive and more resilient cities on the one hand, by "freeing up and using" dormant resources in a largely market driven manner (i.e. no central regulation or control). On the other hand, concerns arise on aspects such as health, equity, discrimination, or distribution outcomes that seem to be simultaneously embedded inherently in these processes, tied deeply to their relationship with the sizes of urban systems (Cottineau, 2016; Sarkar, Phibbs, Simposon and Wasnik, 2016). A natural question arises: as the world urbanises faster than ever, and urban agglomertions grow in number and absolute size, what is the role played by these self-organised peer-to-peer processes? To resolve this dilemma, a deep data-driven understanding is needed. Unfortunately, due to their fast-changing, short term and dynamic nature, they also represent a large data gap in terms of enhancing our understanding of this specific class of socio-economic urban processes and their relationship to spatial structure, since national statistical bodies [such as the Australian Bureau of Statistics (ABS) for Australia] do not have many surveys or programs to capture this type of data.

In line with the major theme of this conference, shared mobility and accommodation practices in cities represent one of the key areas of "big urban data" generation with a very specific characteristic: the data on sharing is big (in the sense of weekly or daily volumes), but it is also big because of its temporal continuity (Santi, 2014). The websites for each of these services witness hourly, daily and weekly changes, with the added caveat that this data is also extremely ephemeral: listings can disappear as soon as a property is let out. Capturing such data can provide a wealth of information and policy relevant evidence for urban planning, especially because it represents a facet of continuous and ephemeral urban data that is not captured at all through the more traditional and official data collection mechanisms. Existing official data collection channels are much more infrequent and dependent on samples, and therefore ill-suited to capture the rich temporal dynamics of such processes. Facilitated by the web footprint, it is now possible to capture very large samples of this class of data, more or less continuously, thereby overriding the problem of small or infrequent samples in traditional surveys.

However, with any new data type, strengths as well as limitations should be discussed. In this case too, there are limitations. First, this type of data does pose quality and reliability questions, which is why the large sample size is important for making any informed analysis. With a large sample size, even with the presence of noise certain statistical trends and regularities may be derived. However, the analysis has to be carefully performed to ensure that the design of analytics is sound. We will discuss this limitation more clearly later in the paper when we analyse rents and geography in the shared accommodation sector. Second, the availability and analysis of this big data does not completely exclude the need for a more human, in-depth examination of the related issues, through more traditional means such as interviews. This human aspect will be especially relevant for understanding issues such as social tie formation, discrimination, overcrowding, health aspects and other such concerns. For example, deep qualitative studies can bring out the more human dimensions, but are restricted by the small sample size to specific areas of a single city (McNamara and Connell, 2007). As research progresses, it will be very useful to develop methodologies that can combine the best of big data-driven approaches, quantitative modelling, and in-depth qualitative explorations.

In this paper, we focus specifically on Internet enabled accommodation sharing web platforms in Australia, that facilitate the phenomenon of flat or house sharing through peer-to-peer direct interactions of those who want to rent with those who want to share or rent out their homes. Studies on this phenomenon are extremely sparse, though there is now evidence on different parts of the world on engaging with different socio-economic questions related to this phenomenon (Kim, 2016; Carlsson, 2015; Ahrentzen, 2003; McNamara and Connell, 2007). We have previously presented a preliminary analysis of this phenomenon in Sydney (Gurran, Phibbs, Sarkar, 2016), and now we extend our analysis to the whole of Australia.

Internet platforms reduce search and transaction costs associated with finding a place to stay (for the new renter), and attracting a lodger or flatmates

(for the existing renter or owner in a property). Platforms such as flatmates.com are flourishing and could radically expand housing choices for renters and home purchasers. However, any evidence on the housing benefits or otherwise of such practices is limited, and there are growing questions about potential risks to public health and safety (e.g. overcrowding), residential amenity, and the supply of affordable rental housing. In this paper, we present a first and preliminary quantitative glance on shared accommodation practices across all major urban centres of Australia enabled via analytics of thousands of online listings. We examine, countrywide, the spatial and short time scale temporal characteristics of the phenomenon, along with preliminary analysis on rents, dwelling types and other dwelling characteristics. Findings have implications for housing policy makers and planning practitioners seeking to monitor and respond to housing policy and affordability pressures in formal and informal housing markets.

## 2. Methodology

We begin by providing a short description of the methodology we followed to collect data on shared accommodation in Australia. We focused on two accommodation sharing sites, gumtree.com.au and flatmates.com.au, that post regular listings for shared homes. We reviewed carefully the legal and privacy notices on both websites, ensuring that no laws are breached in the collection of this data. We also called or emailed representatives from both these organizations. All of these efforts confirmed that the data is provided and accessed freely in the public domain by all parties.

Data was collected by writing Python based scripts that were used weekly, over 8 weeks from October 2016 to December 2016, to collect listings data from both these sites. The same designated day of the week was used to mine this data for all the 6 weeks. 26,864 unique listings were collected for the 5 major urban centres of Australia: Sydney (8 weeks), Melbourne, Brisbane, Perth and Adelaide (all 7 weeks), Figure 1. Due to limitations of space, in this paper we focus only on the data from flatmates.com.au.

We note here that it is not reliable to converge data from the two sites: amongst thousands of listings, we found principally through manually tracking image data and suburb/address data in some specific listings, that there could be double listings. That is, a person could choose to list their place on both gumtree and flatmates, but would be assigned unique ids on each platform. Thus, we have kept the data from the two sources separated.

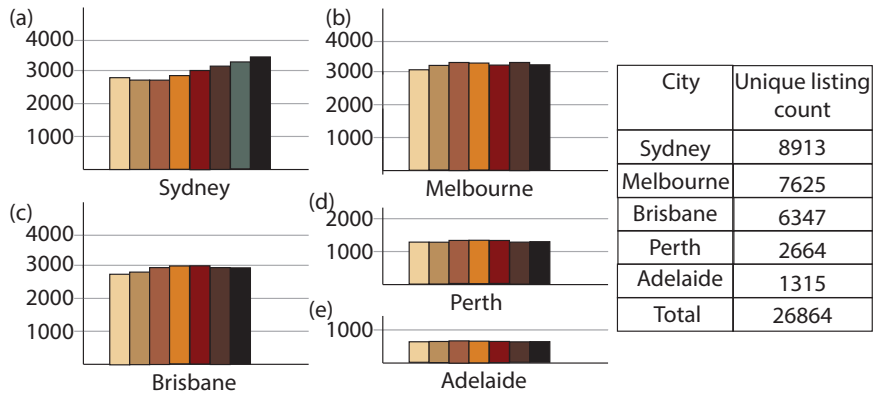

**Figure 1.** Unique weekly listings over 2 months (8 weeks for Sydney and 7 weeks for the other cities) compiled from Flatmates.com. Each coloured bar shows a week's data.

The data was collected in the form of JSON (Java Script Object Notation) variables, and was then parsed and cleaned (to ensure only unique listings are being collected) to produce a final csv (Comma Separated Values) text file, one for each week. Data points that were collected included unique listing ids, city, postcode, suburb, street address, latitude, longitude, dwelling condition information (eg. parking characteristics, bedroom, bathroom and occupancy counts, dwelling type, internet availability, bills included or not included in rent etc.), information on owners, weekly rent, bonds, gender preferences, maximum and minimum number of days of stay, and a text description that provided insight into the demographics of the people renting out the property and the desired demographics of future renters. Matlab and Excel were then used to analyse the data in different ways, the results of which we present in the next section.

## 3. Results

Each of the main results focusses on a target research question that is currently unanswered by traditional data sources. We note here that from the volume of data collected, a wealth of different analytics is possible, but we focus on the most primary of questions in this paper, as exemplars to the variety of possible analytics. All the analysis presented here has been performed for all the 5 major cities, but the results from Sydney are discussed more in a couple of instances, since the highest volume of listings were recorded in Sydney.

### 3.1 What is the spatial spread and geography of the shared accommodation market in Australia?

It is usually assumed that shared accommodation across Australia would be particularly dense around inner city areas and specific areas such as Universities or business districts. However, as Figures 2, 3, and 4, show for Sydney, Melbourne and Brisbane, the shared accommodation market is spread completely over the greater metropolitan regions, albeit getting sparser with distances from the city center. Analysis of Perth and Adelaide show similar patterns. The larger urban regions show evidence of listings in regional areas away from the main city too, e.g. Gosford, Kincumber, Woy Woy in NSW, etc. Although the highest density of listings is recorded around the major CBD areas, newly emerging business districts or educational areas are likely to show a high density of listings as well. For example, the area around Parramatta in Sydney shows a density of listings comparable to Sydney CBD, Figure 3. Particularly important is the dependence of the spatial pattern on the major transit and road routes. For example, in Sydney, the highest numbers of listings occur around Sydney CBD, Eastern Suburbs, and Parramatta, but the trend is most prominent along the major train routes and roads, showing conclusively that transport accessibility is a major determinant of the spatial spread of the shared accommodation market.

A particularly interesting observation can be made around the issue of geographic granularity here. While in our previous analysis of Sydney (Gurran, Phibbs, Sarkar, 2016), we analysed the data at the Local Government Area (LGA) level, in this paper we have adopted a finer granularity and performed the analysis over suburbs and postcodes. We had earlier found that at the LGA level, the City of Sydney and Waverley localities showed the highest density of listings (accounting for 8-14% of all dwellings that were shared homes between non-related adults), while the Parramatta LGA accounted for only 4% of all dwellings accommodating rental sharing. On the other hand, at the suburb level or post code level granularity, it is seen that the suburb of Parramatta alone is in the top 10 suburbs ordered by the number of listings. This is extremely significant since this brings out that there are potentially sharing "hot-spots" that can only be brought out at a finer level of data granularity at the suburb or post code level and that will be missed by coarser LGA level analysis. Further, it also brings out the uneven geography of distribution that may be affecting inner, middle and outer urban areas, where potentially different social and economic "attractors" might be governing the dynamics of the shared market. For example, while it can be expected that areas near universities and holiday hotspots (e.g. Bondi in Sydney) will attract younger home sharers, middle ring shared accommodation patterns (e.g. Parramatta) could instead be shaped by young working families or new migrants who may share because they find rents to be too high or unaffordable at the single-family level. This has potential implications for shared accommodation practices being viewed from two completely different lenses: one from a perspective of the young person transitioning through this particular but temporary phase of housing in their lives, another from a perspective of long term sharing between young, working or migrant families for whom the situation may last several years or become a semi-permanent way of life. Even to explore these hypotheses on long term and short term behaviors and typologies, the starting point is the ready availability of such data and related robust analytic, empirical and modeling mechanisms.

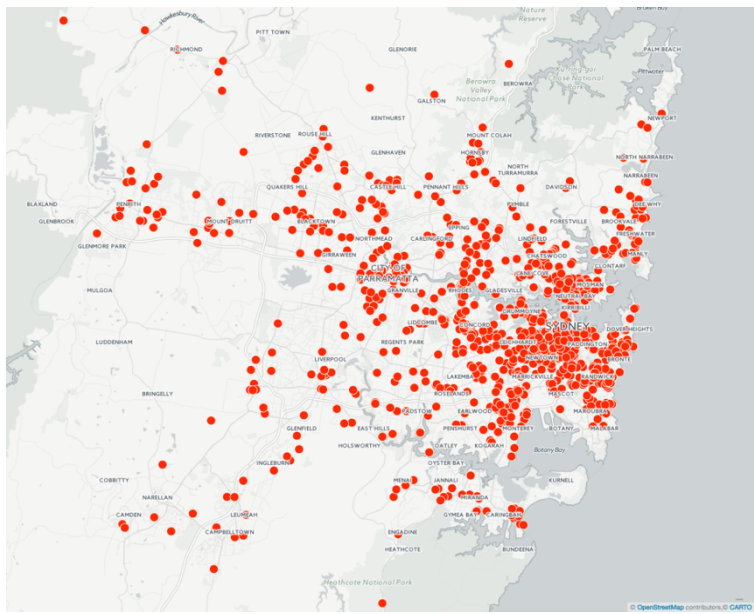

**Figure 2.** Unique listings by latitude-longitude positions on Flatmates.com, Sydney, October 2016.

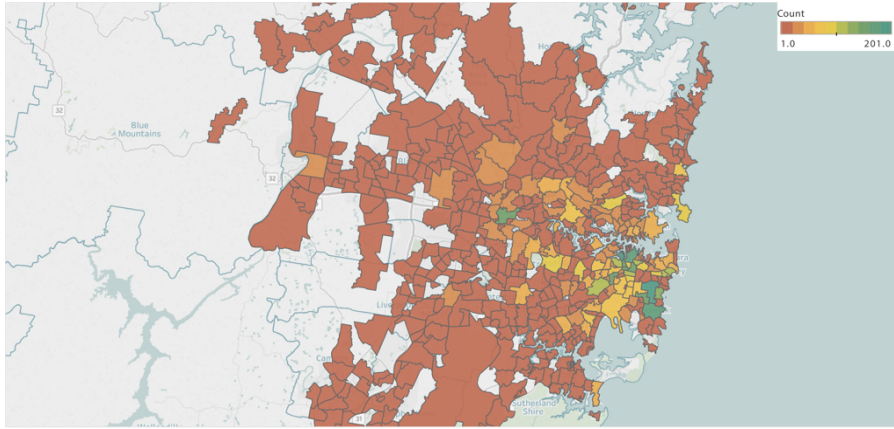

**Figure 3.** Counts of 8913 unique listings on Flatmates.com, Sydney over 8 weeks from October 2016 to December 2016. An interactive version of the map is available at: https://public.tableau.com/shared/HR5FMF9JM?:display_count=yes.

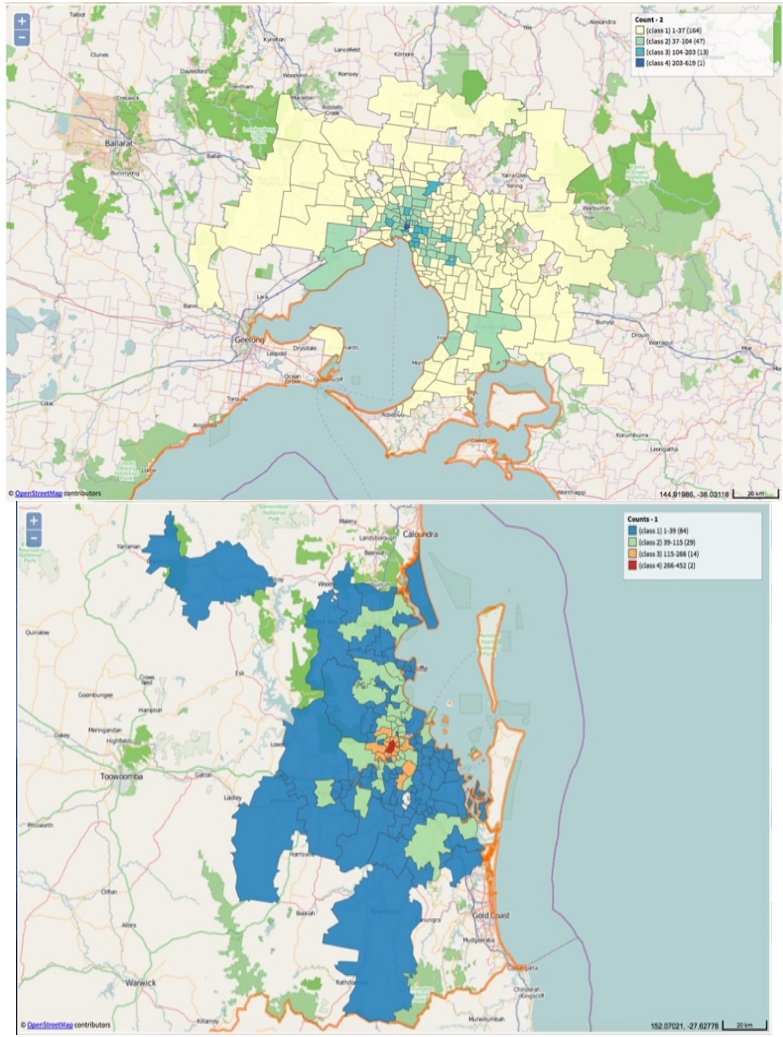

**Figure 4.** Counts of 7625 and 6347 unique listings, respectively, on Flatmates.com, for Melbourne (above) and Brisbane (below) over 7 weeks from November 2016 to December 2016. Melbourne (above) Dark blue: 1 postcode, 3000 (203 - 619 listings), Turquoise blue: 13 postcodes (104-203 listings), Light green: 47 postcodes (37-104 listings), Yellow: 164 postcodes (1-37 listings). Brisbane (below) Red: 2 postcodes (266 - 452 listings), Orange: 14 postcodes (115-266 listings), Green: 29 postcodes (39-115 listings), Blue: 84 postcodes (1-39 listings). Maps produced using AURIN portal.

| Top 10 Sydney suburbs | Unique listing counts over 8 weeks |
|---|---|
| Pyrmont | 201 |
| Randwick | 198 |
| Maroubra | 174 |
| Sydney | 167 |
| Surry Hills | 155 |
| Parramatta | 154 |
| Ultimo | 154 |
| Kingsford | 151 |
| Newtown | 150 |
| Chippendale | 142 |
| Coogee | 140 |

**Table 1.** Top 10 suburbs of Sydney by number of listings.

### 3.2 What are the temporal flow characteristics of the shared accommodation market in Australia?

One particular dimension that could greatly change the perception of the volume occupied by the shared accommodation market is the temporal dimension.

Consider that there are a number of unique listings captured in Week *i*. Three outcomes can occur over time to Week *i+1*: (i) a previous property is rented out, and/or the listing is removed and does not appear in Week *i+1*, (ii) a previous property continues to be on the market, in which case it reappears in the Week *i+1* listings, and (iii) some new properties are advertised in Week *i+1*. An overall analysis of these trends will show the rates of how quickly properties are entering or leaving the market, or the rate of their persistence. 7-8 weeks is too short a time to compute conclusive quantitative numbers like the average number of weeks on the market, or the clearing rate of properties. However, we make a beginning to this type of stock and flow analysis here with our preliminary data. Figure 4 shows the analysis for Sydney, where it is seen that the number of properties entering or leaving each week remains more or less consistent over the short time scale of weeks, though over the 8 weeks a small rise is seen (from 2745 to 3401 listings). Moreover, the number of listings that are removed is roughly equal to the number of listings that are new each week. Further, there seems to be a high degree of persistence of properties on the market. For example, 69% of the listings of Week 1 reappear in Week 2, and 24% of the listings of Week 1 still appear in Week 8. Moreover, the patterns of these percentages are fairly stable, that is, the weekly overlap, the two-weekly overlap, the three-weekly overlap, and so on, all hover around similar percentage values, showing that at least over the short time scale of weeks, the process may be treated as a stationary one. We performed the same analysis for the other 4 cities, and found similar patterns for all of them, with little fluctuation.

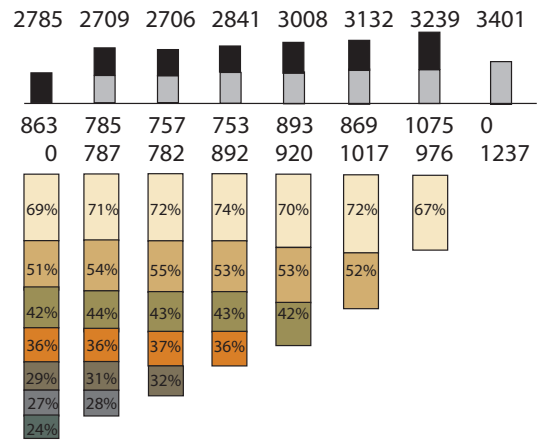

**Figure 4.** Temporal analysis of unique weekly listings for Sydney, 8 weeks from October 2016 to December 2016. Top row of numbers shows total unique listings per week. Second row of numbers, shown as black bars, corresponds to the number of listings that were removed, i.e., number of properties let out that week. Third row of numbers, shown as grey bars, corresponds to the number of listings that were new in that week, i.e., number of properties advertised that week. The coloured bars show the number of listings that persist over the weeks. They are to be read as follows: (i) 69% of 2785 properties in Week 1 continue to be listed in Week 2, 51% of 2785 properties in Week 1 continue to be listed in Week 3, and so on. The time series for listings out weekly, new listings in weekly, and weekly overlaps (that is a measure of the time listings remain on the market) show fairly stable behaviors over the 8 weeks, and therefore the process may be treated as a stationary one in short time scales of weeks or months.

### 3.3 What are the dwelling characteristics and typologies of the shared accommodation market in Australia?

In the formal housing market, housing a larger number of people has a direct correlation to the production of new dwellings. In the shared accommodation or the informal sector on the other hand, it is the existing dwelling stock that "expands" to accommodate a higher number of people. Thus, the dwelling typologies and distributions appeared to be governed largely by the physical characteristics of the existing dwelling stocks in the 5 cities. Figure 5 shows the breakdown of the dwelling types across the 5 cities. Shared flats and shared houses come out to be the most prominent categories, but other typologies are present in all of the cities (e.g., one bedroom properties, studios, granny flats, etc.). "Whole property" forms a separate class because these are primarily shared accommodation platforms, and thus, it is rarer for entire properties to be advertised here.

There is a subtle difference between the patterns observed in Sydney versus the other cities. In Sydney, flatshare forms the dominant mode of sharing (51% of listings) while housesharing comes at a close second (48% of listings). In all the other cities, the trend is opposite: housesharing forms the most prominent form of sharing with flatsharing the second most prominent. It is worth noting that in Perth and Adelaide househaring is the only prominent mode. Figure 6 shows the same data but broken down by the number of bedrooms. Again, there is a prominent trend: 2 and 3 bedroom properties form the largest components of the shared housing market. Thus, the size of the city seems to have a direct correlation with the available dwelling stock typologies, and is therefore shaping behavior in the shared accommodation market.

Since one of the main concerns raised for the shared accommodation market is the possibility of overcrowding, we compared the number of bedrooms with the occupancy count variables of the listings. We counted all dwellings where the occupant count was listed as higher than the number of bedrooms, under the assumption that this could serve as a measure of potential overcrowding. Table 2 shows the results. Again, the size of the city seems to play a central role, with Sydney showing the largest rates of potential overcrowding, and Adelaide the least. In a few individual examples, it would appear from the listing data that as many as 6 to 7 occupants are sharing dwellings with 2 or 3 bedrooms, although there are also a small number of examples where 6 bedroom dwellings have upto 7 occupants.

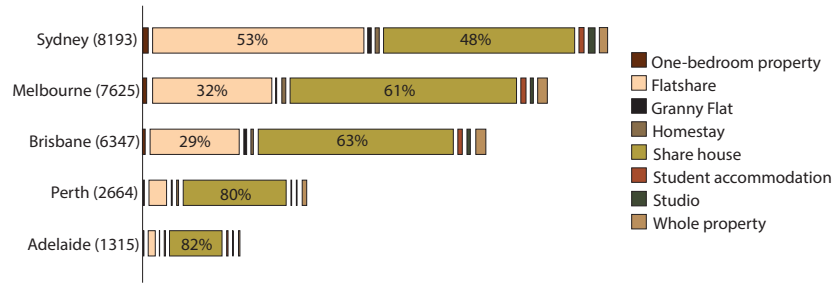

**Figure 5.** Dwelling Types Analysis: 26864 unique listings across 5 major urban centres

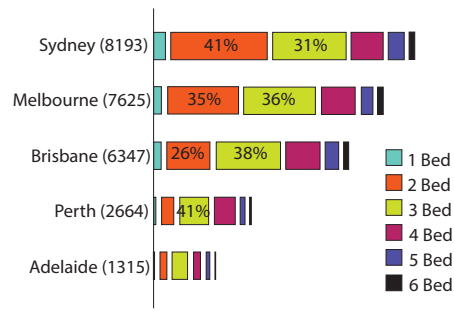

**Figure 6.** Bedroom Type Distribution: 26864 unique listings across 5 major urban centres

| City | Actual number of listings where occupant count is higher than the number of bedrooms | Percentage of the total number of listings (Note: in computing total number of listings, all those listings where the occupant count information was not available and had entry: "None" were excluded) |
|---|---|---|
| Sydney | 945 out of 7478 | 13% |
| Melbourne | 580 out of 6420 | 9% |
| Brisbane | 385 out of 5553 | 7% |
| Perth | 93 out of 2228 | 4% |
| Adelaide | 48 out of 1130 | 4% |

Table 2. Occupant Count versus No. of Bedrooms analysis to predict potential overcrowding

### 3.4 What are the rental characteristics of the shared accommodation market in Australia?

The primary hypothesis that needs to be examined with regard to rental trends is *whether the presence of shared accommodation in an area can actually push up the local rents for individual (i.e. non-shared) renting, and at the same time increase the risk of overcrowding in the locality*. There is sufficient motivation for this hypothesis. First, the localities that are the most popular and have the highest access characteristics to local education and employment opportunities will attract the largest number of shared properties. Second, these could the very localities where investors are motivated to purchase the most, with the result, that when the owner of a property is not living in the dwelling (i.e., the property is an investment property), they will have incentive to encourage overcrowding to maximize the rents that can be gained from a single property. Indeed, in our dataset, an extremely high proportion of the dataset recorded the "owner" category as "false", implying that either most people in these listings are likely to be a renter rather than owner, or that these properties are managed by agents or investor owners. Thus, a detailed and careful framework of analysis would need to be developed to ensure that the analysis of rents is sound, and not based on simplifying assumptions, and produces a correct estimate of rents per dwelling in the area based on the data. This is likely a topic of future research, given the complexity of the data, but we have made some preliminary analysis here through a case study on Pyrmont in Sydney as an exemplar, the suburb with the highest number of listings. We note here, as we have before, that middle and outer ring suburbs could have driving dynamics that are different from the inner suburbs.

For example, consider the following quote from one of the listings from Pyrmont, Sydney, for a 2 bedroom flat being shared by 7 people, and looking for the $8^{th}$ person (noting that there are multiple entries of this "type", i.e. this case is not a unique exception but presents an example of a class with several similar database entries):

ONLY FEMALE UNIT~~NO MINIMUM STAY~!Masterbedroom 4persons room $135 per week. Second room 4 persons room $125 per week- Very friendly family-like girls~- People are from Australia German French Korea China Japan- Very Clean and spacious

As per realestate.com.au (accessed December 2016 – March 2017), the current median rent for 2 bedroom dwellings in Pyrmont is $775/week, for 3 bedroom dwellings is $990/week, and for 4 bedrooms is $2000/week. If there are 8 people in a 2 bedroom dwelling with the rates mentioned above,

this implies a total weekly rent of $1040/week, which is far higher than the median of $775/week.

Thus, in carrying out a rent analysis, the actual distribution of the number of bedrooms and the number of occupants per dwelling at a fine geographic granularity would need to be accurately considered. For example, the average number of bedrooms in Pyrmont listings, over all the two and three bedroom listings that form the dominant share, was computed to be 2, whereas the average occupant count came to 3. Thus, the rent per room per dwelling would need to be tripled on average to get an approximate idea of the average rent being gained from a 2 bedroom property in Pyrmont. Figure 7 shows a tree map with the rent data for the top 20 Sydney suburbs. For Pyrmont, the average weekly per room rent for all 2 and 3 bedroom properties was computed as $300, which shows an average of about $900/week for a 2 bedroom property, as compared to the $775/week for a 2 bedroom property via individual non-shared renting.

A further observation can be made regarding the geographic distribution of rents from Figure 7. It is seen that the volume of listings (i.e. a measure of "supply") in a suburb does not correlate as highly with the average and median rents (i.e., a measure of "price"), as does the distance from the city center. For the case of Sydney, Figure 7 shows that the suburbs located in the inner city and the eastern suburbs show the highest rents, whereas rents go down progressively for the high volume middle income suburbs.

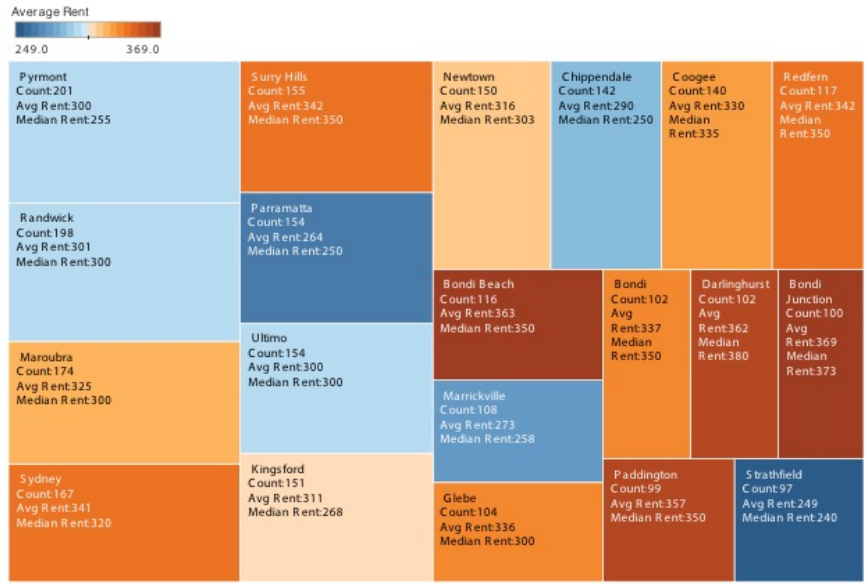

**Figure 7.** Average and Median rents for 2 and 3 bedroom shared accommodation in the top 20 Sydney suburbs by listing counts. In the treemap, the size of a box corresponds to the number of listings, and the colour corresponds to the average rent. There seems to be no correlation between number of listings and rents, instead, the distance from the CBD emerges as a principal factor: rents are the highest around the CBD and the eastern suburbs, and gradually go down towards the west and southwest of of Sydney.

## 4. Conclusions and Discussion

In this paper, we studied the "data" opportunity provided by the online shared accommodation market in Australia, across its 5 major urban centres. We reported on the methodology used for data collection, and presented a preliminary analysis of the spatial, temporal and structural characteristics of this market as revealed by the data. We explored four specific research questions related to (a) the geography, (b) the temporal flow characteristics, and (c) the dwelling characteristics, and (d) the rental characteristics of the shared accommodation market in Australia.

Shared economies and shared urbanism are self-organised phenomena. The most pertinent theoretical question that arises from the investigation is whether these processes have the capacity to make cities more adaptive, responsive and resilient, or whether, they also simultaneously work to increase inequitable distributions of resources in cities that get spatially embedded over time as a result of these processes. While it would be too early to answer this question now, the hope is that the facility of having access to more and more datasets of this nature will trigger new patterns of research and methodology development, that could help us answer questions of resilience, adaptive and responsive urban dynamics in the future.